\documentclass[amssymb,aps,twocolumn,floats]{revtex4}
\usepackage{color,graphicx,pstricks}
\draft
\begin{document}

\title{ Spin-spiral states in undoped manganites }

\author{Sanjeev Kumar$^{1,2}$, Jeroen van den Brink$^{1,3,4,5}$, and Arno P. Kampf$^{6}$}

\affiliation{
$^{1}$ Institute Lorentz for Theoretical Physics, Leiden University,
P.O. Box 9506, 2300 RA Leiden, The Netherlands \\
$^{2}$ Faculty of Science and Technology,
University of Twente, P.O. Box 217, 7500 AE Enschede, The Netherlands \\
$^{3}$ Institute for Molecules and Materials, Radboud Universiteit Nijmegen,
P.O. Box 9010, 6500 GL Nijmegen, The Netherlands \\
$^{4}$ Stanford Institute for Materials and Energy Sciences, Stanford University and SLAC National Accelerator Laboratory, 
Menlo Park, CA 94025, USA \\
$^{5}$ Leibniz-Institute for Solid State and Materials Research Dresden, D-01171 Dresden, Germany \\
$^{6}$ Theoretical Physics III, Electronic Correlations and Magnetism,
University of Augsburg, D-86135 Augsburg, Germany
}

\begin{abstract}
The experimental observation of multiferroic behavior in perovskite manganites 
with a spiral spin structure demands to clarify the origin of these magnetic 
states and their relation to ferroelectricity. We show that spin-spiral phases 
with diagonal wavevector and also the E-type phase exist for intermediate 
values of the Hund's rule and the Jahn-Teller coupling in the orbitally ordered
and insulating state of the standard two-band model Hamiltonian for manganites.
Our results support the spin-current mechanism for ferroelectricity and present
an alternative view to earlier conclusions where frustrating superexchange 
couplings were crucial to obtain spin-spiral states.

\vskip 0.1cm


\end{abstract}

\maketitle

The coexistence of long-range magnetic order with spontaneous electric 
polarization is commonly refered to as multiferroic behavior 
\cite{khomskii_review}. In recent years multiferroic materials have attracted 
special attention from the condensed matter community because of their 
potential for applications in memory and data storage devices 
\cite{ramesh,ahn}. Despite the initial observation that materials with 
coexisting ferroelectric and magnetic orders are rare in nature \cite{spaldin},
an increasing number of multiferroic materials with interesting properties has
been discovered \cite{kimura_nature,hur,lawes,kimura_PRL}. Different mechanisms
for the origin of multiferroic behavior have been proposed and partially 
identified \cite{dag2, picozzi, spin_current, mostovoy, JvdB, CO}. In a 
selected class of multiferroics ferroelectricity is driven by the existence of 
a non-trivial magnetic order, e.g., with a spiral spin structure 
\cite{kim,seki,kimura,goto}. It has therefore become a key issue to clarify 
the conditions under which non-trivial spin states can occur in different 
models and materials.

Hole-doped perovskite manganites are known for their rich phase diagrams and 
complex transport phenomena \cite{dagotto_book}. The recent observations of 
ferroelectricity in TbMnO$_3$ and DyMnO$_3$ stimulated further research also 
on the undoped materials. The magnetic groundstate of RMnO$_3$ with R = La,
Pr, Nd and Sm is an A-type antiferromagnet (AFM), with ferromagnetic (FM) 
order in the a-b plane and a staggered spin pattern along the c-axis. It 
changes to a spiral magnet for R = Tb and Dy, and finally to an E-type AFM for
R = Ho, where zigzag FM chains alternate in their preferred spin direction 
\cite{kimura,kimura1}. Although the A-type order of the prototype compound 
LaMnO$_3$ is well understood in terms of Goodenough-Kanamori rules and orbital 
ordering \cite{dagotto_book}, the magnetic structure of the materials with 
smaller ionic radii are much less understood. It was proposed that due to 
GdFeO$_3$-type structural distortions longer range interactions become relevant
and thereby may lead to complex magnetic groundstates with spiral or E-type 
spin patterns \cite{kimura, dagotto_spiral}. The existence of an E-type 
pattern in these models with longer range interactions has been put into 
question recently \cite{comment}.

In this letter we explore the magnetic groundstates in a two-band model for 
RMnO$_3$ without invoking the next-nearest neighbor or even longer range 
interactions. The model consists of itinerant electrons coupled locally to 
core spins via the Hund's rule coupling $J_H$, and a nearest neighbor 
antiferromagnetic interaction between the spins. Importantly, we refrain from 
taking the commonly adopted double-exchange limit $J_H \rightarrow \infty$. The
two-orbital nature of the model allows for the inclusion of the Jahn-Teller 
(JT) distortions as a source for orbital order and hence the insulating 
character of the undoped manganites. We find that both the spiral spin states 
and the E-type states are stable in the parameter regime relevant for 
manganites. The experimentally observed transitions between different magnetic
states are obtained only for finite JT couplings and intermediate values of 
$J_H$.

Specifically, we consider a two-dimensional two-band model with the 
Hamiltonian \cite{JvdB_Khomskii}:
\begin{eqnarray}
H &=& -\sum_{\langle ij \rangle \sigma}^{\alpha \beta}
t_{\alpha \beta} \left ( c^{\dagger}_{i \alpha \sigma} c^{~}_{j \beta \sigma} 
+ H.c. \right ) \cr
&&
 - J_H \sum_{i \alpha} {\bf S}_i \cdot {\mbox {\boldmath $\sigma$}}_{i \alpha}
+ J_s \sum_{\langle ij \rangle} {\bf S}_i \cdot {\bf S}_j ~ ~ ~ ~ ~ ~ ~ ~
\end{eqnarray}
\noindent
at quarter filling to describe the undoped manganites RMnO$_3$. Here, 
$c^{}_{i\alpha\sigma}$ and $c^{\dagger}_{i\alpha\sigma}$ are annihilation and 
creation operators for electrons with spin $\sigma$ in the $e_g$ orbital 
$\alpha \in \{x^2-y^2, 3z^2-r^2\}$, which from here onwards is labeled as $1$ 
and $2$, respectively. $t_{\alpha\beta}^{ij}$ denote the hopping amplitudes 
between the two $e_g$ orbitals on nearest-neighbor sites and have the cubic 
perovskite specific form \cite{dagotto_book}: $t_{11}^x= t_{11}^y \equiv t$,
$t_{22}^x= t_{22}^y \equiv t/3 $, $t_{12}^x= t_{21}^x \equiv -t/\sqrt{3} $,
$t_{12}^y= t_{21}^y \equiv t/\sqrt{3} $, where $x$ and $y$ mark the spatial 
directions. In a commonly used approximation for manganites, the core spins 
are treated as classical vectors with $|{\bf S}|=1$; the justification of this 
approximation was quantitatively verified \cite{neuber}.
${\mbox {\boldmath $\sigma$}}_{i \alpha}$ denotes the electronic spin operator 
defined as
${\sigma}^{\mu}_{i \alpha}=
\sum_{\sigma \sigma'}^{\alpha} c^{\dagger}_{i\alpha \sigma}
\Gamma^{\mu}_{\sigma \sigma'}
c_{i \alpha \sigma'}$,
where $\Gamma^{\mu}$ are the Pauli matrices. The one-band version of the above 
model and the Kondo lattice model with $J_s=0$ have been previously analyzed 
in one- and two-dimensions in search for non-trivial magnetic groundstates 
\cite{hamada,hallberg, nandini, pinaki}. Although spin-spiral states were 
found for selected combinations of band fillings and Hund's rule coupling, the
connection to the spin spirals observed in the RMnO$_3$ is unclear since the
insulating character of orbitally ordered RMnO$_3$ is not captured by the 
one-band models.

By applying a canonical transformation we rewrite the Hamiltonian in a basis 
where the spin-quantization axis is site-dependent and points along the 
direction of the local core spin. Introducing polar and azimuthal angles 
$\theta$ and $\phi$, respectively, the transformation is defined as:
\begin{eqnarray}
\left[ \begin{array}{c}
c_{i \alpha \uparrow} \\
c_{i \alpha \downarrow}
\end{array} \right] & = &
\left[ \begin{array}{c c} \cos \left( \frac{\theta_i}{2} \right)~ e^{{\rm i} 
\phi_i/2} &
- \sin \left( \frac{\theta_i}{2} \right)~ e^{{\rm i} \phi_i/2} \\
\sin \left( \frac{\theta_i}{2} \right )~ e^{-{\rm i}\phi_i/2} & \cos \left( 
\frac{\theta_i}{2} \right)~ e^{-{\rm i}\phi_i/2}
\end{array} \right]
\left[ \begin{array}{c}
d_{i \alpha p} \\
d_{i \alpha a}
\end{array} \right] \nonumber  \\
& \equiv & {\cal U}(\theta_i, \phi_i)
\left[ \begin{array}{c}
d_{i \alpha p} \\
d_{i \alpha a}
\end{array} \right]
\end{eqnarray}
\noindent Here $d^{}_{i \alpha p}$ ($d^{}_{i \alpha a}$) annihilates an 
electron at site $i$ in orbital $\alpha$ with spin parallel (antiparallel) to 
the core spin. In terms of d operators the Hamiltonian reads:
\begin{eqnarray}
H &=& -\sum_{\langle ij \rangle \sigma}^{\alpha \beta} \sum_{s,s'}
t_{\alpha \beta} \left ( f_{ss'} d^{\dagger}_{i \alpha s} d^{~}_{j \beta s'} + 
H.c. \right ) \cr
&&
 -\frac{J_H}{2} \sum_{i \alpha} (n_{i \alpha p} - n_{i \alpha a})
 + J_s \sum_{\langle ij \rangle} {\bf S}_i \cdot {\bf S}_j ~, ~ ~ ~
\end{eqnarray}
\noindent
with $n_{i \alpha s} = d^{\dagger}_{i \alpha s} d^{~}_{i \alpha s} $. The 
coefficients $f_{ss'}$, are explicitly given by,
\begin{eqnarray}
\left[ \begin{array}{c c} f_{pp} & f_{pa} \\
f_{ap} & f_{aa} \end{array} \right] & = & {\cal U}^{\dagger}(\theta_i, \phi_i).
{\cal U}(\theta_j, \phi_j),
\end{eqnarray}
\noindent
The advantage of this transformation is that the Hund's rule term now becomes 
diagonal. For a fixed configuration of classical spins the Hamiltonian is 
bilinear in the fermion operators. However the one-particle Schr\"odinger 
equation can not be solved in closed form for an arbitrary spin configuration. 
We therefore selectively analyze the core spin configurations described by the 
polar and azimuthal angles $\theta_i = \Theta$ and $\phi_i = {\bf q} \cdot 
{\bf r}_i $. We refer to $\Theta$ as the cone angle and ${\bf q}$ as the 
spiral wavevector. Despite this restriction most of the well known spin 
patterns as observed in various magnetic materials are included, e.g., 
ferromagnetic, antiferromagnetic, canted-ferromagnetic, and spin-spiral 
patterns.

For this choice of variational spin states the Hamiltonian matrix reduces to a $4 \times 4$ matrix
after Fourier transformation with, $d^{}_{i \alpha s} = N^{-1/2}\sum_{\bf k} e^{ -{\rm i} {\bf k} \cdot {\bf r}_i} ~ d^{}_{{\bf k} \alpha s}$,
and the kinetic energy term in Eq. (3) is then written as, $H_{kin} = \sum_{{\bf k}} {\bf D}^{\dagger}_{\bf k} ~ {\cal H} ({\bf k}) ~ {\bf D}^{}_{\bf k}$,
where, $ {\bf D}^{\dagger}_{\bf k} \equiv \left [ d^{\dagger}_{{\bf k} 1 p} ~ d^{\dagger}_{{\bf k} 1 a} ~ d^{\dagger}_{{\bf k} 2 p}
~ d^{\dagger}_{{\bf k} 2 a} \right ]$, and the $4 \times 4$ matrix ${\cal H}$ is given by
\begin{eqnarray}
{\cal H}({\bf k}) \equiv \left [ \begin{array}{c c c c}
h^{pp}_{11} & h^{pa}_{11} & h^{pp}_{12} & h^{pa}_{12} \\
h^{ap}_{11} & h^{aa}_{11} & h^{ap}_{12} & h^{aa}_{12} \\
h^{pp}_{21} & h^{pa}_{21} & h^{pp}_{22} & h^{pa}_{22} \\
h^{ap}_{21} & h^{aa}_{21} & h^{ap}_{22} & h^{aa}_{22}
\end{array} \right ],
\end{eqnarray}
\noindent
with the matrix elements,
\begin{eqnarray}
h^{pp}_{\alpha \beta} & = & 2\sum_{\mu = x,y} t^{\mu}_{\alpha \beta}[ \cos^2 
\left( \frac{\Theta}{2} \right) \cos k^{+}_{\mu}
+ \sin^2 \left( \frac{\Theta}{2} \right) \cos k^{-}_{\mu} ] \nonumber \\
h^{aa}_{\alpha \beta} & = & 2\sum_{\mu = x,y} t^{\mu}_{\alpha \beta} [ \cos^2 
\left( \frac{\Theta}{2} \right) \cos k^{-}_{\mu}
+ \sin^2 \left( \frac{\Theta}{2} \right) \cos k^{+}_{\mu} ] \nonumber \\
h^{ap}_{\alpha \beta} & = & 2\sum_{\mu = x,y} t^{\mu}_{\alpha \beta} 
\sin(\Theta) \sin(\frac{q_{\mu}}{2}) \sin k_{\mu} ~ 
= ~ h^{pa}_{\alpha \beta},
\end{eqnarray}
\noindent
with $k^{\pm}_{\mu} = k_{\mu} \pm q_{\mu}/2$.
For a single band $H_{kin}$ reduces to a $2 \times 2$ matrix structure and therefore the eigenspectrum is straightforwardly obtained in a closed form expression \cite{hamada}. For the two-band case considered here the closed form result for the dispersion is rather involved and we therefore diagonalize the above $4 \times 4$ matrix numerically for each momentum ${\bf k}$ on finite lattices of upto $256 \times 256$ sites.

\begin{figure}[t]
\centerline{
\includegraphics[width=8.8cm , clip=true]{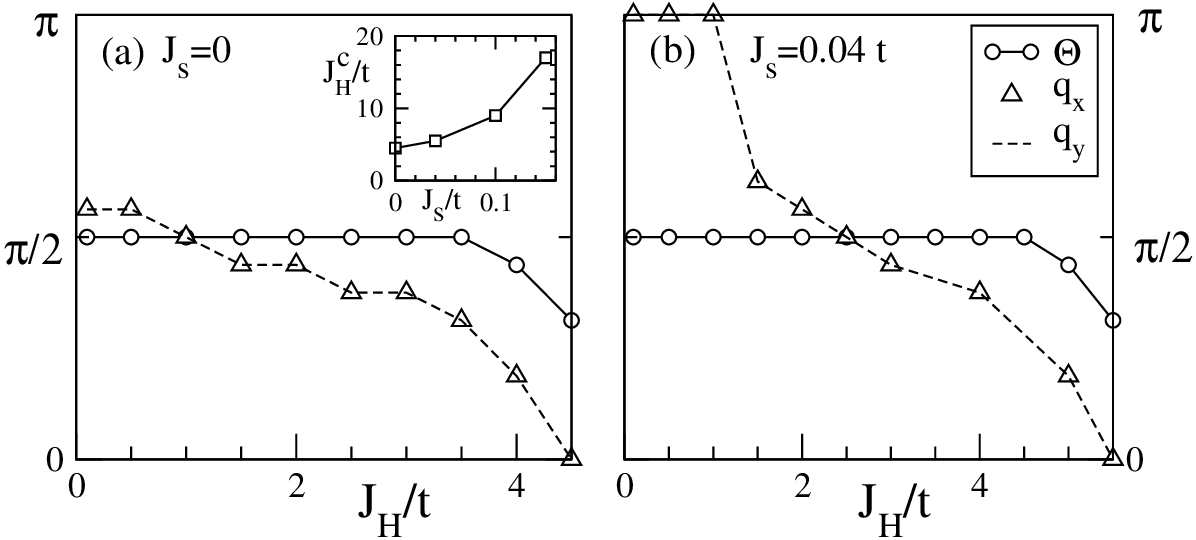}
}
\vspace{.1cm}
\caption{~ The cone angle $\Theta$ and the wavevector $(q_x,q_y)$ of the lowest-energy spiral state as a function of $J_H$ for, (a) $J_s = 0$ and (b) $J_s = 0.04 t$. The inset in (a) shows the $J_s$ dependence of the critical value of $J_H$ required to obtain the FM state.
}
\end{figure}

In Fig. 1 we plot the values of the cone angle $\Theta$ and spiral wavevector $(q_x,q_y)$
corresponding to the minimum-energy state as a function of $J_H$.
Over almost entire parameter regime the cone angle is found to be $\pi/2$, which corresponds to planer spin states. Another feature is that $q_x = q_y$, suggesting that the diagonal spirals are more stable, which is consistent with the experimental findings \cite{kimura1}. For the diagonal spirals with $(q_x,q_y) = q(1,1)$ the spiral pitch $q$ smoothly vanishes upon reaching the FM state with $q=0$. Close to the FM phase the cone angle of the spiral state slightly deviates from $\pi/2$. The groundstate jumps discontinuously from an antiferromagnet with $q = \pi$ to a spiral with $q < \pi$. Inset in Fig. 1(a) shows the variation with $J_s$ of the critical value of Hund's rule coupling $J^c_H$ for the transition to a ferromagnet.

Before arriving at the magnetic phase diagram of the Hamiltonian, it is essential to consider other states which are not captured by the ansatz ($\theta_i,\phi_i$)=($\Theta,{\bf q}\cdot{\bf r}_i$) \cite{hallberg, aliaga}.
The energies of the earlier suggested candidate states, including the E-type states, for the magnetic order in manganites are
obtained by exact numerical diagonalization and compared with those of the spiral states.

The groundstate-phase diagram for the quarter filled system is shown in Fig. 2(a). The E-type phase is stable in a wide region of parameter space. Spiral states are favored for larger values of $J_s$, and also in a narrow window between
the FM and the E-type states. Here, the FM state in a two-dimensional model is representative of a single plane of the A-type AFM state. In the experiments on RMnO$_3$ two transitions are observed, first from the A-type AFM to the spiral state and subsequently to the E-type phase, upon reducing the ionic radius of the rare-earth element \cite{kimura}. Therefore the existence of a direct FM to E-type transition appears to contradict the experimental observation.
Moreover, the phase diagram of Fig. 2(a) corresponds to a metallic state with a finite density of states at the Fermi level, the undoped manganites however are insulating. Therefore it is essential to include the source for the opening of an energy gap in the spectrum
in order to obtain results applicable to RMnO$_3$. For this purpose we add an adiabatic Jahn-Teller coupling to the Hamiltonian, which is given by,
\begin{eqnarray}
H_{JT} = \lambda \sum_{i} [Q_{x i}\tau_{x i}+Q_{z i}\tau_{z i}]+\frac{K}{2} 
\sum_i |{\bf Q}_i|^2.
\end{eqnarray}
\noindent
In Eq. (7), $Q_{x i}$ and $Q_{z i}$ are lattice distortions corresponding to 
two different JT modes.
$\tau_{x i} = \sum_{\sigma} (c^{\dagger}_{i 1 \sigma} c^{~}_{i 2 \sigma} +
c^{\dagger}_{i 2 \sigma} c^{~}_{i 1 \sigma}) $ and
$\tau_{z i} = \sum_{\sigma} (c^{\dagger}_{i 1 \sigma} c^{~}_{i 1 \sigma} -
c^{\dagger}_{i 2 \sigma} c^{~}_{i 2 \sigma}) $
are orbital pseudospin operators \cite{dagotto_book}. In the undoped manganites
staggered JT distortions are accompanied by orbital ordering with transition 
tempertatures much higher than the temperatures scale for magnetic ordering 
\cite{kimura}. Therefore the pattern for the JT distortions is expected to be 
robust upon cooling even though the magnitude of the distortions may depend on 
the magnetic structure. A real-space Monte Carlo study verified the staggered 
ordering of the $Q_x$ component when $H_{JT}$ is included in the Hamiltonian 
\cite{sk-apk-pm}. Therefore we adopt this pattern for the lattice distortions 
and set $Q_{xi}=Q^0_{x}~e^{{\rm i}(\pi,\pi)\cdot{\bf r}_i}$ and $Q_{zi} = 0 $.
The second term in Eq. (7) is the energy cost associated with the distortion 
of the lattice with $|{\bf Q}_i|^2=Q_{xi}^2+Q_{zi}^2$. We set the stiffness 
constant $K=t$ as in previous theoretical model analyses of manganites 
\cite{dagotto_book,sk-apk-pm}. We treat $Q^0_x$ as a
variational parameter in the calculations and optimize it by minimizing the 
total energy. Therefore $Q^0_x$ is allowed to vary in the different magnetic 
states.

Due to the staggered orbital order the determination of the eigenspectrum becomes slightly more involved. Instead of a $4 \times 4$ matrix, we now have to diagonalize the $ 8 \times 8$ matrix
$
\left [ \begin{array}{c c}
{\cal H}({\bf k}) & {\cal M} \\
{\cal M} & {\cal H}({\bf k}')
\end{array} \right ]
$ for each ${\bf k}$ with ${\bf k}' = {\bf k} + (\pi,\pi)$ and
\begin{eqnarray}
{\cal M} = \left [ \begin{array}{c c c c}
0 & 0 & \lambda Q^0_x & 0 \\
0 & 0 & 0 & \lambda Q^0_x \\
\lambda Q^0_x & 0 & 0 & 0 \\
0 & \lambda Q^0_x & 0 & 0
\end{array} \right ].
\end{eqnarray}
\noindent
The energy minimization procedure with respect to $\Theta$ and ${\bf q}$ is followed as before but in the presence of staggered Jahn-Teller distortions, which lead to an insulating state with staggered orbital order. The resulting phase diagram is shown in Fig. 2(b). Remarkably the inclusion of the Jahn-Teller distortions leads to the appearance of spiral states in between the E-type and the FM phases in a wide range of the Hund's rule coupling.

\begin{figure}[t]
\centerline{
\includegraphics[width=8.8cm , clip=true]{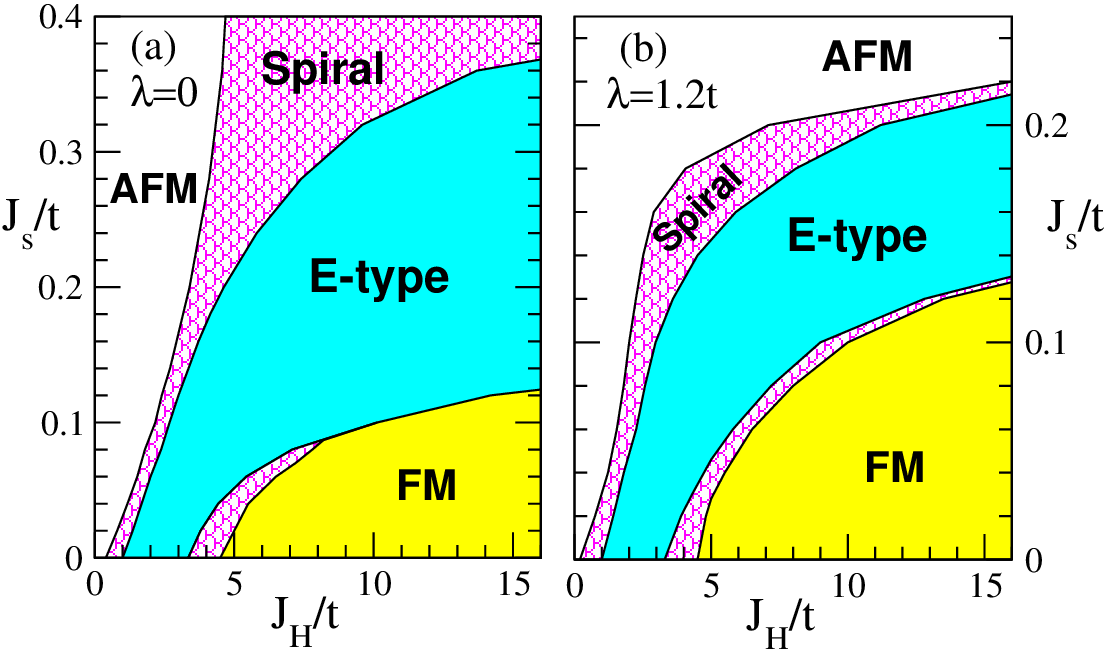}
}
\vspace{.1cm}
\caption{~ (Color online) Groundstate-phase diagrams $J_s$ versus $J_H$ in the absence (a) and presence (b) of Jahn-Teller distortions. The spiral states have the cone angle $\Theta = \pi/2$ except in the narrow region between FM and the E-type states in (a). In both cases $q_x = q_y$ reduces monotonically as the FM state is approached.
}
\end{figure}

An important effect of the size reduction of the rare earth ion in RMnO$_3$ is 
the decrease of the electronic bandwidth due to the Mn-O-Mn bond angle moving 
further away from $180^{\circ}$. For the model calculations, it is simpler to 
vary $J_s$ and $\lambda$ rather than changing the hopping parameters which 
control the bandwidth. In Fig. 3 we therefore show the phase diagrams in the 
parameter space of $J_s$ and $\lambda$ for two representative values of $J_H$.
For $J_H = 5 t$, the small $J_s$ regime is ferromagnetic, which describes a 
single plane of the A-type AFM observed in RMnO$_3$ with R = La, Pr, Nd and Sm. A two-step
transition occurs from FM via the spiral to the E-type state by increasing 
both $\lambda$ and $J_s$ (indicated by the arrow in Fig. 3(a)), which 
effectively translates to reducing the bandwidth. In the $(1,1)$ spiral state 
the pitch $q$ increases along the direction of the arrow (see Fig. 3(a)). The 
values of $q$ at the two end-points of the planar spiral phase along the arrow
are $0.13 \pi$ and $0.25 \pi$. These values are close to the experimental
results of $0.14 \pi$ and $0.19 \pi$ obtained for TbMnO$_3$ and DyMnO$_3$, 
respectively \cite{kimura1}. The pitch vector ${\bf q'} = (0,k,0)$ reported in
the experiments translates to ${\bf q} = (k/2,k/2,0)$ on the Mn square lattice 
used here due to a 45$^{\circ}$ rotation between the two coordinate systems.
The E-type AFM state has been observed in HoMnO$_3$ \cite{kimura,kimura1}.

For the larger Hund's rule coupling $J_H$ the stability region of the spiral states shrinks considerably (see Fig. 3(b)). Moreover, the spiral states are no longer sandwiched between the FM and the E-type phases. This implies that it is not possible in the commonly adopted double-exchange ($J_H \rightarrow \infty$) limit to find transitions to spiral states as observed in RMnO$_3$ upon varying the bandwidth, unless further interactions are added to the two-band model Hamiltonian.

\begin{figure}[t]
\centerline{
\includegraphics[width=8.8cm , clip=true]{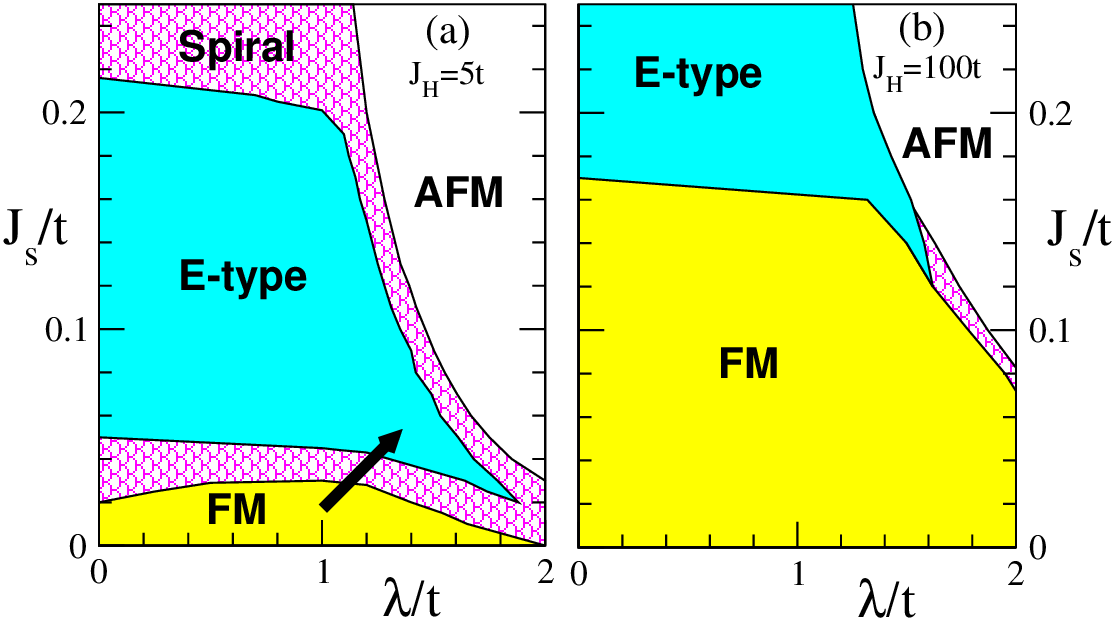}
}
\vspace{.1cm}
\caption{~ (Color online) $J_s$-$\lambda$ phase diagrams for (a) $J_H = 5t$, and (b) $J_H=100t$.
The arrow in panel (a) is indicative of the path traced upon reducing the ionic radius of the rare-earth ion in RMnO$_3$, which leads to the correct sequence of transitions from FM to spiral to E-type states. Such a sequence can not be traced along any straight line in panel (b).
}
\end{figure}

The presence of a spiral structure in an insulator has been identified as one 
possible source for a spontaneous electric polarization ${\bf P}$ by generating
spin currents \cite{spin_current}. The direction of ${\bf P}$ is perpendicular 
to both the direction of the spiral pitch vector ${\bf q}$ and the cone axis of
the spiral \cite{mostovoy}. Our model of choice is isotropic in spin space and
thus can not determine the orientation of the cone axis relative to the
crystallographic directions.
Using the input from the experiments regarding the direction of the cone axis
for the spiral state we indeed obtain the direction of  ${\bf P}$ consistent with
the experiments \cite{kimura1, yamasaki}.
Within the spin-current mechanism the magnitude of {\bf P} is controlled by the length of
the pitch vector $q$, which we obtained in the experimentally relevant range.

Therefore our results verify that already the standard two-band model for
manganites has all the necessary ingredients to sustain the magnetic spiral and
the E-type phases as observed in the undoped perovskite manganites. A finite
Hund's rule coupling of the order of the bandwidth leads to a spiral pattern
and a wavelength which both compare well with the observed magnetic structure 
in TbMnO$_3$ and DyMnO$_3$. These results support the applicability of the 
spin current mechanism as the source for ferroelectricity in RMnO$_3$. Lattice
distortions of the GdFeO$_3$-type are likely to give rise to additional
longer-range couplings, which may further stabilize the spiral and the E-type
states. The essential physics of these non-trivial spin states is already
contained in the simpler short-range two-orbital model with finite $J_H$.

We thank E. Dagotto for useful comments. This work was supported by NanoNed, 
FOM, and by the Deutsche Forschungsgemeinschaft through SFB 484.

{}

\end{document}